**Magnetic anisotropy dependence of the energy of oriented thick ferromagnetic films**

P. Samarasekara

Department of Physics, University of Ruhuna, Matara, Sri Lanka

**ABSTRACT**

Heisenberg Hamiltonian was employed to describe the variation of energy of thick ferromagnetic films with second and fourth order anisotropies. At $\theta=1.36^0$ and $\frac{\sum_{m=1}^{N} D_m^{(4)}}{\omega} = \frac{\sum_{m=1}^{N} D_m^{(2)}}{\omega} = 1.25$, energy is minimum for thick film of sc(001) with 1000 layers. Energy becomes minimum at $\theta=1.18^0$ and $\frac{\sum_{m=1}^{N} D_m^{(4)}}{\omega} = 1.15$ for thick film of bcc(001) with the same thickness. According to these simulations, these lattices can be easily oriented in some certain directions under the influence of some particular values of anisotropies. Energy varies with $\frac{\sum_{m=1}^{N} D_m^{(4)}}{\omega}$ and $\frac{\sum_{m=1}^{N} D_m^{(2)}}{\omega}$ in similar passion for both types of lattices. The energy gradually decreases with $\frac{\sum_{m=1}^{N} D_m^{(4)}}{\omega}$ and $\frac{\sum_{m=1}^{N} D_m^{(2)}}{\omega}$ for both types of lattices in the range of $\frac{\sum_{m=1}^{N} D_m^{(4)}}{\omega}$ and $\frac{\sum_{m=1}^{N} D_m^{(2)}}{\omega}$ described here.

**1. INTRODUCTION:**

Properties of ferromagnetic films receive a wide attention nowadays, due to their potential applications of magnetic memory devices and microwave devices [1]. Because of the difficulties of understanding the behavior of exchange anisotropy and its applications in magnetic sensors and media technology, the theoretical investigations of exchange anisotropy have been extensively expanded [2]. Bloch spin wave theory is sometimes applied to study magnetic properties of ferromagnetic thin films [3]. The magnetization of some thin films is oriented in plane due to dipole interaction. But the perpendicular orientation is preferred at the surface due to the broken symmetry of uniaxial anisotropy energy. Two dimensional Heisenberg model has been earlier used to explain the magnetic anisotropy in the presence of dipole interaction [4]. Magnetic properties of ferromagnetic thin films with alternating super layers have been thoroughly investigated by Ising model [5].

The stress induced anisotropy is very important in the investigations of soft ferromagnetic materials, since the stress induced anisotropy is in the range of the crystal anisotropy of soft ferromagnetic materials [6-11]. Therefore, the stress induced anisotropy has been taken into account for the studies given here. Under the influence of an applied magnetic field, the demagnetization effect takes place inside a magnetic thin film due to the separation of magnetic poles. But the dipole interaction takes place even without the applied field. On the other hand, the dipole interaction is a microscopic parameter, and demagnetization factor is a macroscopic parameter. Therefore, the both dipole interaction and the demagnetization factor would be taken into account to study the behavior of these thick films completely. Heisenberg Hamiltonian was used to investigate the ferrite films [10, 12, 13, 14, 16, 17] and ferromagnetic films [11, 15].

## 2. MODEL AND DISCUSSION:

The classical Heisenberg Hamiltonian of ferromagnetic thin films can be given in following form [10-11].

$$H = -\frac{J}{2}\sum_{m,n}\vec{S}_m \cdot \vec{S}_n + \frac{\omega}{2}\sum_{m \neq n}\left(\frac{\vec{S}_m \cdot \vec{S}_n}{r_{mn}^3} - \frac{3(\vec{S}_m \cdot \vec{r}_{mn})(\vec{r}_{mn} \cdot \vec{S}_n)}{r_{mn}^5}\right) - \sum_m D_{\lambda_m}^{(2)}(S_m^z)^2 - \sum_m D_{\lambda_m}^{(4)}(S_m^z)^4$$
$$- \sum_{m,n}[\vec{H} - (N_d \vec{S}_n / \mu_0)] \cdot \vec{S}_m - \sum_m K_s \sin 2\theta_m \qquad (1)$$

Here m and n represent indices of two different layers, N is the number of layers measured in direction perpendicular to the film plane, J is the magnetic spin exchange interaction, $Z_{|m-n|}$ is the number of nearest spin neighbors, ω is the strength of long range dipole interaction, $\Phi_{|m-n|}$ are constants for partial summation of dipole interaction, $D_m^{(2)}$ and $D_m^{(4)}$ are second and fourth order anisotropy constants, $H_{in}$ and $H_{out}$ are components of applied magnetic field in plane and out of plane of the film, $N_d$ is the demagnetization factor, and $K_s$ is the constant related to the stress which depends on the magnetization and the magnitude of stress.

The energy of oriented ferromagnetic thick films can be given as following for discrete case,

$$E(\theta) = -\frac{J}{2}[NZ_0 + 2(N-1)Z_1]$$

$$+ \{N\Phi_0 + 2(N-1)\Phi_1 + 2ae^{-2b}[\frac{N-2+e^{-b}(1-N)}{(1-e^{-b})^2}]\}(\frac{\omega}{8} + \frac{3\omega}{8}\cos 2\theta)$$

$$-\cos^2\theta \sum_{m=1}^{N} D_m^{(2)} - \cos^4\theta \sum_{m=1}^{N} D_m^{(4)} - N(H_{in}\sin\theta + H_{out}\cos\theta - \frac{N_d}{\mu_0} + K_s \sin 2\theta) \quad (2)$$

For sc(001) lattice with $Z_0=4$, $Z_1=1$, $\Phi_0=9.0336$, $\Phi_1=-0.3275$, $\frac{J}{\omega}=1.06 \times 10^3$, $a=-16\pi^2$, $b=2\pi$ and $N=1000$,

$E(\theta)= -3179 \times 10^3 \omega + 1047 \omega(1 + 3\cos 2\theta)$

$$-\cos^2\theta \sum_{m=1}^{N} D_m^{(2)} - \cos^4\theta \sum_{m=1}^{N} D_m^{(4)} - 1000(H_{in}\sin\theta + H_{out}\cos\theta - \frac{N_d}{\mu_0} + K_s \sin 2\theta)$$

When $\frac{N_d}{\mu_0 \omega} = 10$, $\frac{\sum_{m=1}^{N} D_m^{(2)}}{\omega} = 4000$, $\frac{H_{out}}{\omega} = 5$, $\frac{H_{in}}{\omega} = 5$, and $\frac{K_s}{\omega} = 5$

$$\frac{E(\theta)}{\omega} = -3168 \times 10^3 + 3141 \cos 2\theta - 4000\cos^2\theta - \frac{\sum_{m=1}^{N} D_m^{(4)}}{\omega} \cos^4\theta$$

$$-1000(5\cos\theta + 5\sin\theta + 5\sin 2\theta)$$

The 3-D plot of energy versus angle and $\frac{\sum_{m=1}^{N} D_m^{(4)}}{\omega}$ is given in figure 1. The minimum of energy can be observed at $\theta=1.36^0$ and $\frac{\sum_{m=1}^{N} D_m^{(4)}}{\omega}=1.25$ implying that the thick film with 1000 layers can be easily oriented in that direction under the influence of this particular fourth order anisotropy constant. The anisotropy constant can be different from layer to layer, and here 1.25 is the addition of all the anisotropy constants of all the layers.

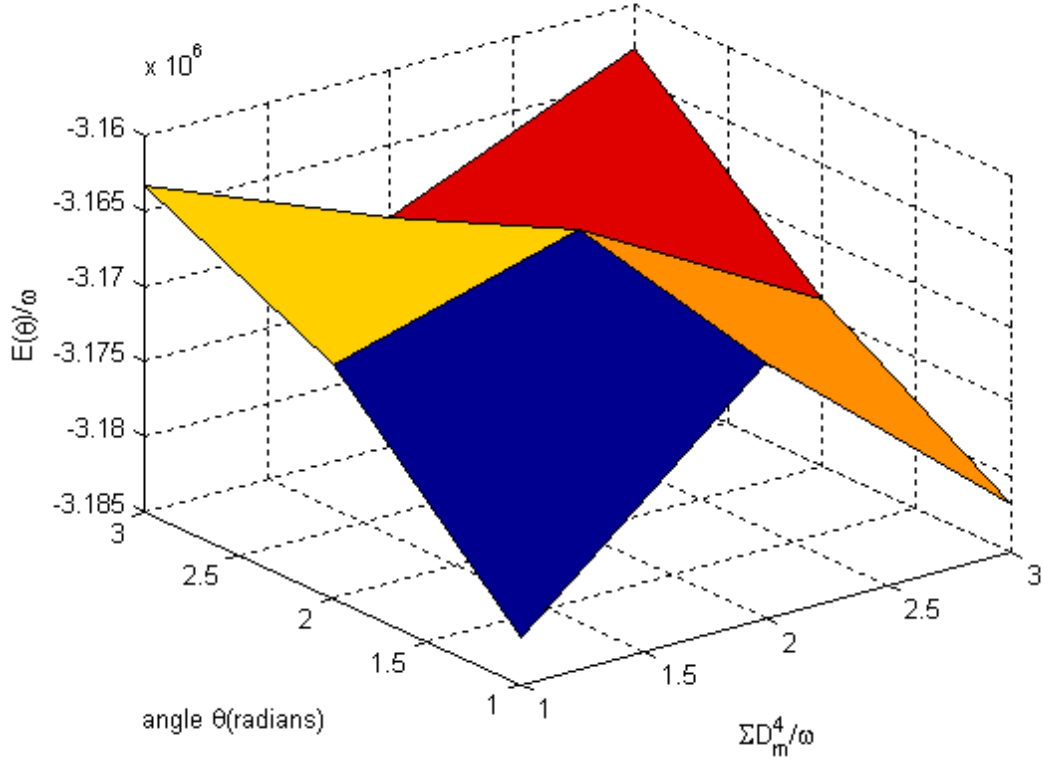

Figure 1: 3-D plot of energy versus angle and $\dfrac{\sum_{m=1}^{N} D_m^{(4)}}{\omega}$ for sc(001) with N=1000

When $\dfrac{N_d}{\mu_0 \omega} = 10$, $\dfrac{\sum_{m=1}^{N} D_m^{(4)}}{\omega} = 189$, $\dfrac{H_{out}}{\omega} = 5$, $\dfrac{H_{in}}{\omega} = 5$, and $\dfrac{K_s}{\omega} = 5$

$$\dfrac{E(\theta)}{\omega} = -3168 \times 10^3 + 3141 \cos 2\theta - \dfrac{\sum_{m=1}^{N} D_m^{(2)}}{\omega} \cos^2\theta - 189\cos^4\theta$$

$$-1000(5\cos\theta + 5\sin\theta + 5\sin 2\theta)$$

The 3-D graph of energy versus angle and $\frac{\sum_{m=1}^{N} D_m^{(2)}}{\omega}$ is given in figure 2. Again the energy is minimum at θ=1.36⁰ and $\frac{\sum_{m=1}^{N} D_m^{(2)}}{\omega} = 1.25$. $\frac{\sum_{m=1}^{N} D_m^{(4)}}{\omega}$ and $\frac{\sum_{m=1}^{N} D_m^{(2)}}{\omega}$ behave in the same way according to graphs 1 and 2. But the energy in this case is larger than the energy in previous case.

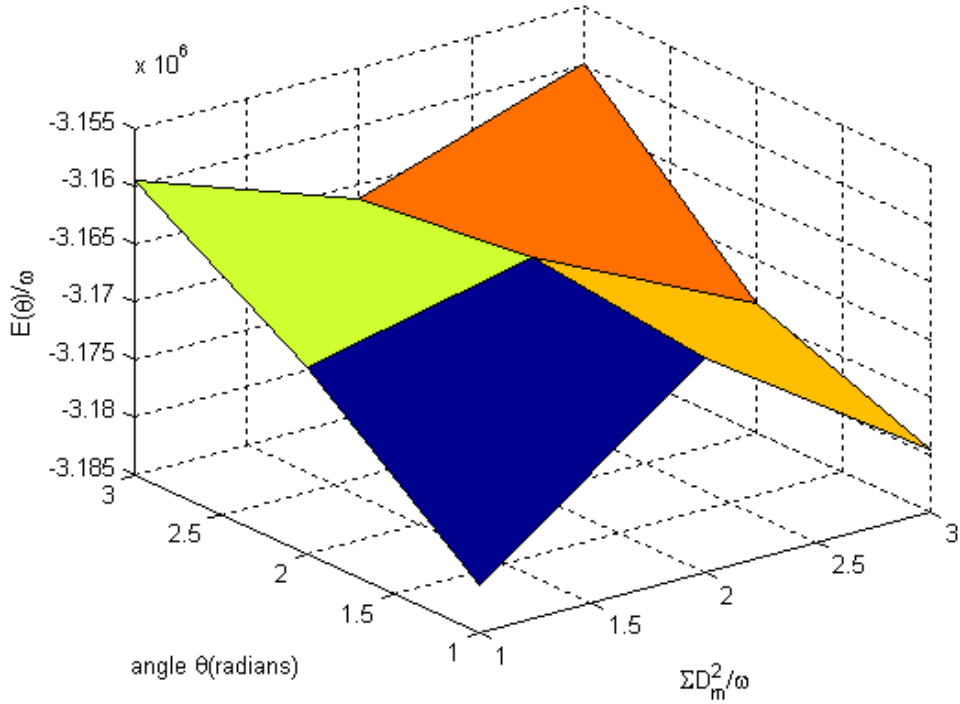

Figure 2: 3-D graph of energy versus angle and $\frac{\sum_{m=1}^{N} D_m^{(2)}}{\omega}$ for sc(001) with N=1000

When θ=0, from above equation for sc(001) lattice

$$\frac{E(\theta)}{\omega} = -3170 \times 10^3 - \sum_{m=1}^{N} D_m^{(2)} - \sum_{m=1}^{N} D_m^{(4)}$$

When the energy is zero, addition of two anisotropy constants is negative according to above equation. Because the anisotropy constant of some layers can be negative, the addition of all anisotropy constants can be also negative. The 3-D plot of energy versus $\dfrac{\sum_{m=1}^{N} D_m^{(2)}}{\omega}$ and $\dfrac{\sum_{m=1}^{N} D_m^{(4)}}{\omega}$ is given in figure 3. The energy gradually decreases with anisotropy constants, and the energy related to any value of anisotropy can be determined using this 3-D plot.

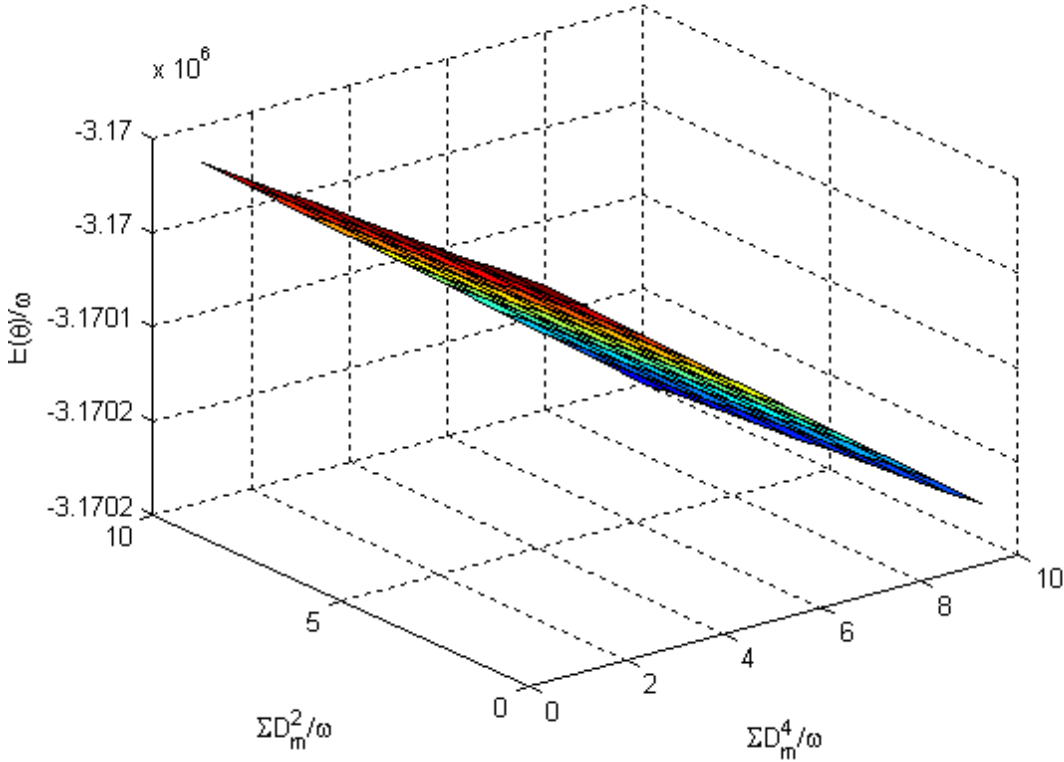

Figure 3: 3-D plot of energy versus $\dfrac{\sum_{m=1}^{N} D_m^{(2)}}{\omega}$ and $\dfrac{\sum_{m=1}^{N} D_m^{(4)}}{\omega}$ for sc(001) with N=1000

Using $Z_0=0$, $Z_1=4$, $\Phi_0=5.8675$, $\Phi_1=2.7126$, $b=\pi$ and $a = 6\sqrt{3}\pi^2$ for bcc(001) lattice in equation number 2,

$E(\theta)= -4240 \times 10^3 \omega + 1465\, \omega(1+3\cos 2\theta)$

$-\cos^2\theta \sum_{m=1}^{N} D_m^{(2)} - \cos^4\theta \sum_{m=1}^{N} D_m^{(4)} - 1000(H_{in}\sin\theta + H_{out}\cos\theta - \dfrac{N_d}{\mu_0} + K_s \sin 2\theta)$

When $\dfrac{N_d}{\mu_0 \omega} = 10$, $\dfrac{\sum_{m=1}^{N} D_m^{(2)}}{\omega} = 4000$, $\dfrac{H_{out}}{\omega} = 5$, $\dfrac{H_{in}}{\omega} = 5$, and $\dfrac{K_s}{\omega} = 5$

$$\dfrac{E(\theta)}{\omega} = -4229 \times 10^3 + 4395 \cos 2\theta - 4000 \cos^2 \theta - \dfrac{\sum_{m=1}^{N} D_m^{(4)}}{\omega} \cos^4 \theta$$

$$-1000(5\cos\theta + +5\sin\theta + 5\sin 2\theta)$$

The 3-D graph of energy versus angle and $\dfrac{\sum_{m=1}^{N} D_m^{(4)}}{\omega}$ is given in figure 4. According to this graph, energy is smaller than the energy of sc(001) graph. But the variation of energy is similar to that of sc(001) lattice. Energy is minimum at $\theta = 1.18^0$ and $\dfrac{\sum_{m=1}^{N} D_m^{(4)}}{\omega} = 1.15$, indicating that the film can be easily oriented along $\theta = 1.18^0$ direction under anisotropy $\dfrac{\sum_{m=1}^{N} D_m^{(4)}}{\omega} = 1.15$.

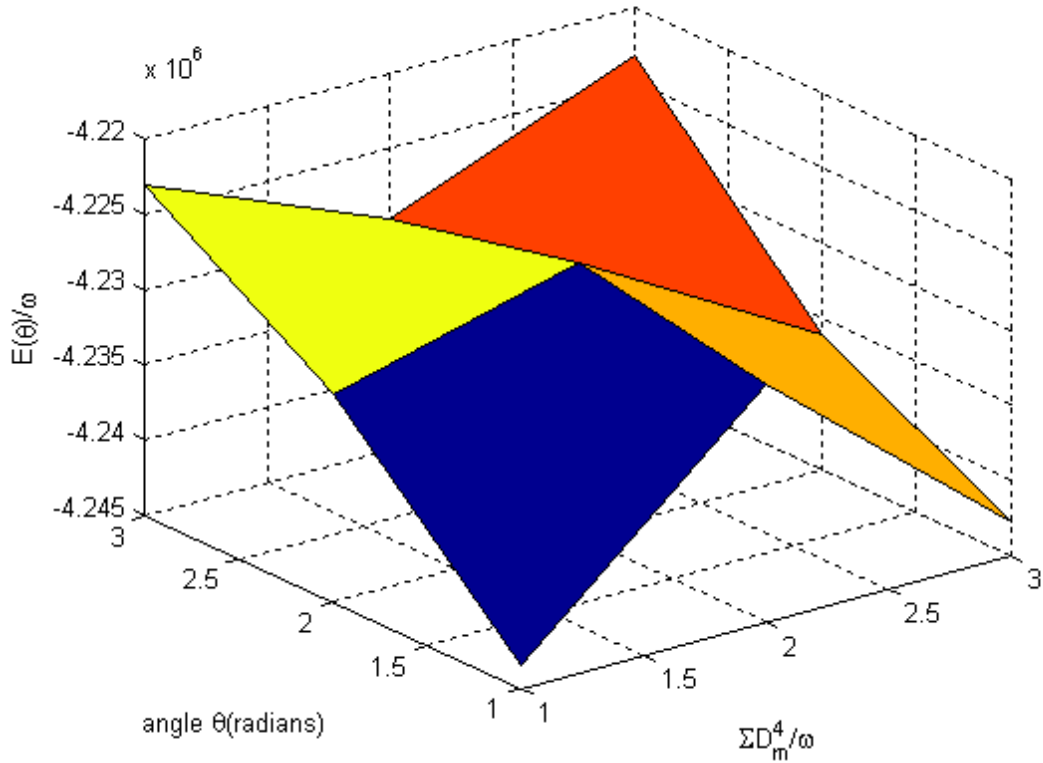

Figure 4: 3-D graph of energy versus angle and $\dfrac{\sum_{m=1}^{N} D_m^{(4)}}{\omega}$ for bcc(001) with N=1000

When θ=0, from above equation for bcc(001) lattice

$$\dfrac{E(\theta)}{\omega} = -4230 \times 10^3 - \sum_{m=1}^{N} D_m^{(2)} - \sum_{m=1}^{N} D_m^{(4)}$$

The 3-D plot of energy versus $\dfrac{\sum_{m=1}^{N} D_m^{(2)}}{\omega}$ and $\dfrac{\sum_{m=1}^{N} D_m^{(4)}}{\omega}$ is given in figure 5. Energy in this case is smaller than that of sc(001) lattice. But the variation of energy is similar to that of sc(001) lattice.

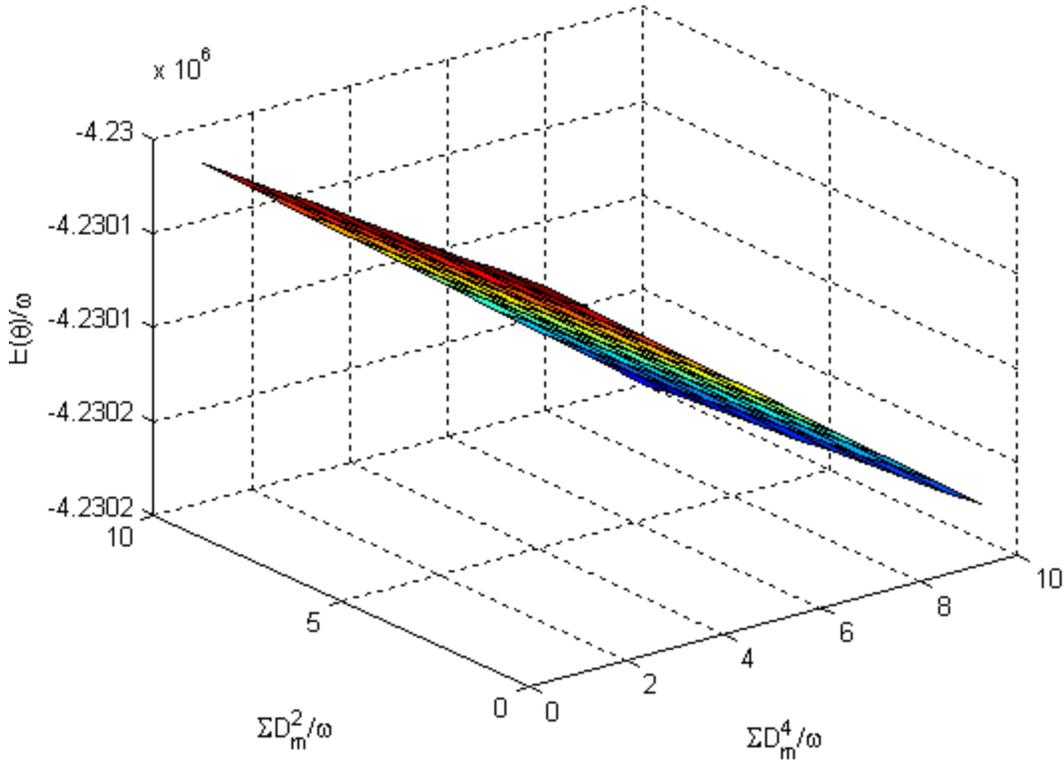

Figure 5: 3-D plot of energy versus $\dfrac{\sum_{m=1}^{N} D_m^{(2)}}{\omega}$ and $\dfrac{\sum_{m=1}^{N} D_m^{(4)}}{\omega}$ for bcc(001) with N=1000

## 3. CONCLUSION:

The energy becomes minimum at $\theta=1.36^0$ and $\dfrac{\sum_{m=1}^{N} D_m^{(4)}}{\omega} = \dfrac{\sum_{m=1}^{N} D_m^{(2)}}{\omega} = 1.25$ for thick film of sc(001) with 1000 layers. Energy is minimum at $\theta=1.18^0$ and $\dfrac{\sum_{m=1}^{N} D_m^{(4)}}{\omega} = 1.15$ for thick film of bcc(001) with 1000 layers. This implies that these lattices can be easily oriented in these certain directions under the influence of these particular values of anisotropies. Variation of energy with $\dfrac{\sum_{m=1}^{N} D_m^{(4)}}{\omega}$ and $\dfrac{\sum_{m=1}^{N} D_m^{(2)}}{\omega}$ is similar in all cases described here. The energy gradually decreases with $\frac{\sum_{m=1}^{N} D_m^{(4)}}{\omega}$ and $\frac{\sum_{m=1}^{N} D_m^{(2)}}{\omega}$ for both types of lattices. But the energy of bcc(001) is smaller than that of sc(001) for all these cases. Although this simulation was performed for some selected values of $\frac{J}{\omega}, \frac{N_d}{\mu_0 \omega}, \frac{\sum_{m=1}^{N} D_m^{(2)}}{\omega}, \frac{\sum_{m=1}^{N} D_m^{(4)}}{\omega}, \frac{H_{out}}{\omega}, \frac{H_{in}}{\omega}$ and $\frac{K_s}{\omega}$ only, this same simulation can be carried out for any other vales as well.

**REFERENCES:**

1. K.D. Usadel and A. Hucht, 2002. Anisotropy of ultrathin ferromagnetic films and the spin reorientation transition. Physical Review B 66, 024419-1.
2. David Lederman, Ricardo Ramirez and Miguel Kiwi, 2004. Monte Carlo simulations of exchange bias of ferromagnetic thin films on $FeF_2$ (110). Physical Review B 70(18), 184422.
3. Martin J. Klein and Robert S. Smith, 1951. Thin ferromagnetic films. Physical Review 81(3), 378-380.
4. M. Dantziger, B. Glinsmann, S. Scheffler, B. Zimmermann and P.J. Jensen, 2002. In-plane dipole coupling anisotropy of a square ferromagnetic Heisenberg monolayer. Physical Review B 66(9), 094416.
5. M. Bentaleb, N. El Aouad and M. Saber, 2002. Magnetic properties of the spin -1/2 Ising Ferromagnetic thin films with alternating superlattice configuration. Chinese Journal of Physics 40(3), 307.
6. P. Samarasekara and F.J. Cadieu, 2001. Polycrystalline Ni ferrite films deposited by RF sputtering techniques. Japanese Journal of Applied Physics 40, 3176-3179.
7. P. Samarasekara and F.J. Cadieu, 2001. Magnetic and Structural Properties of RF Sputtered Polycrystalline Lithium Mixed Ferrimagnetic Films. Chinese Journal of Physics 39(6), 635-640.

8. H. Hegde, P. Samarasekara and F.J. Cadieu, 1994. Nonepitaxial sputter synthesis of aligned strontium hexaferrites, SrO.6($Fe_2O_3$), films. Journal of Applied Physics 75(10), 6640-6642.

9. P. Samarasekara, 2003. A pulsed rf sputtering method for obtaining higher deposition rates. Chinese Journal of Physics 41(1), 70-74.

10. P. Samarasekara, 2007. Classical Heisenberg Hamiltonian solution of oriented spinel ferrimagnetic thin films. Electronic Journal of Theoretical Physics 4(15), 187-200.

11. P. Samarasekara and S.N.P. De Silva, 2007. Heisenberg Hamiltonian solution of thick ferromagnetic films with second order perturbation. Chinese Journal of Physics 45(2-I), 142-150.

12. P. Samarasekara, 2010. Determination of energy of thick spinel ferrite films using Heisenberg Hamiltonian with second order perturbation. Georgian electronic scientific journals: Physics 1(3), 46-49.

13. P. Samarasekara, 2011. Investigation of Third Order Perturbed Heisenberg Hamiltonian of Thick Spinel Ferrite Films. Inventi Rapid: Algorithm Journal 2(1), 1-3.

14. P. Samarasekara and William A. Mendoza, 2011. Third order perturbed Heisenberg Hamiltonian of spinel ferrite ultra-thin films. Georgian electronic scientific journals: Physics 1(5), 15-18.

15. P. Samarasekara and William A. Mendoza, 2010. Effect of third order perturbation on Heisenberg Hamiltonian for non-oriented ultra-thin ferromagnetic films. Electronic Journal of Theoretical Physics 7(24), 197-210.

16. P. Samarasekara, M.K. Abeyratne and S. Dehipawalage, 2009. Heisenberg Hamiltonian with Second Order Perturbation for Spinel Ferrite Thin Films. Electronic Journal of Theoretical Physics 6(20), 345-356.

17. P. Samarasekara and Udara Saparamadu, 2013. Easy axis orientation of Barium hexa-ferrite films as explained by spin reorientation. Georgian electronic scientific journals: Physics 1(9), 10-15.